\documentstyle[titlepage,11pt]{article}

\makeatletter%
\newtheorem{theorem}{Theorem}[section]

\newcommand{\singlespacing}{\let\CS=
\@currsize\renewcommand{\baselinestretch}{1}\tiny\CS}
\newcommand{\singlespacingplus}{\let\CS=
\@currsize\renewcommand{\baselinestretch}{1.25}\tiny\CS}
\newcommand{\doublespacing}{\let\CS=
\@currsize\renewcommand{\baselinestretch}{1.75}\tiny\CS}
\newcommand{\draftspacing}{\let\CS=
\@currsize\renewcommand{\baselinestretch}{2.0}\tiny\CS}

\makeatother%
\newcommand{\np}{{\rm NP}}
\newcommand{\p}{{\rm P}}
\newcommand{\DP}{{\rm DP}}
\newcommand{\implies}{\:\Rightarrow\:}
\newcommand{\conp}{{\rm coNP}}
\newcommand{\SAT}{{\mbox{\tt Sat}}}
\newcommand{\OMS}{\mbox{{\tt OddMaxSat}}}

\newtheorem{definition}[theorem]{Definition}

\makeatletter
\newcommand{\niceonespacing}{\let\CS=\@currsize\renewcommand{\baselinestretch}{1.1}\tiny\CS}\newcommand{\nicetwospacing}{\let\CS=\@currsize\renewcommand{\baselinestretch}{1.2}\tiny\CS}
\newcommand{\nicethreespacing}{\let\CS=\@currsize\renewcommand{\baselinestretch}{1.3}\tiny\CS}
\newcommand{\singlespacingplusplus}{\let\CS=\@currsize\renewcommand{\baselinestretch}{1.35}\tiny\CS}
\newcommand{\nicefourspacing}{\let\CS=\@currsize\renewcommand{\baselinestretch}{1.4}\tiny\CS}
\newcommand{\nicefivespacing}{\let\CS=\@currsize\renewcommand{\baselinestretch}{1.5}\tiny\CS}
\newcommand{\nicesixspacing}{\let\CS=\@currsize\renewcommand{\baselinestretch}{1.6}\tiny\CS}
\makeatother

\makeatletter
\def\@cite#1#2{[#1\if@tempswa , #2\fi]}
\makeatother
\makeatletter
\def\@citex[#1]#2{\if@filesw\immediate\write\@auxout{\string\citation{#2}}\fi
  \def\@citea{}\@cite{\@for\@citeb:=#2\do
    {\@citea\def\@citea{,\linebreak[0]}\@ifundefined
       {b@\@citeb}{{\bf ?}\@warning
       {Citation `\@citeb' on page \thepage \space undefined}}%
\hbox{\csname b@\@citeb\endcsname}}}{#1}}
\makeatother

\title{On the Power of Positive Turing Reductions}

\author{ 
{\em  Edith Hemaspaandra\/}\thanks{  Research supported 
in part by the 
National Science Foundation under grants
NSF-INT-9513368/DAAD-315-PRO-fo-ab and 
NSF-INT-9815095/\protect\linebreak[0]DAAD-315-PPP-g\"u-ab.
Work done in part while visiting
Friedrich-Schiller-Universit\"at Jena.  Email: eh@cs.rit.edu.}
\\
Department of Computer Science\\
Rochester Institute of Technology\\
Rochester, NY 14623}

\date{May 5, 1999}

\begin{document}

    { \singlespacing
      \maketitle
    }

\begin{abstract}

\singlespacing 

In the early 1980s, Selman's seminal work on positive Turing reductions 
showed that positive Turing reduction to NP yields no greater 
computational power than NP itself.  Thus, positive Turing and 
Turing reducibility to NP differ sharply unless the polynomial
hierarchy collapses.

We show that the situation is quite different for DP,
the next level of the boolean hierarchy.  In particular, 
positive Turing reduction to DP already yields all (and only) 
sets Turing
reducibility to NP\@.  Thus, positive Turing and Turing reducibility to DP
yield the same class.   Additionally, we show that an even 
weaker class, $\p^{\np[1]}$, can be substituted for DP in this 
context.

\end{abstract}

\section{Background and Definitions}
A quarter century ago, Selman initiated the study of
polynomial-time positive Turing reductions.  
A truth-table version of this 
reducibility had been introduced a few years
earlier, by Ladner, Lynch, and Selman~\cite{lad-lyn-sel:j:com}.
Polynomial-time positive Turing reductions are defined as
follows.  

Let $\Sigma$ will be any fixed alphabet having at least two
letters.  For specificity, in this paper 
we will take $\Sigma = \{0,1\}$, but that is
not essential.  For any machine $M$, 
$L(M)$ denotes the set 
of strings accepted by machine $M$, and 
for any set $A$,
$L(M^A)$ denotes the set 
of strings accepted by machine $M$ running with oracle $A$.
$A \leq_T^p B$ exactly if there is a polynomial-time Turing 
machine $M$ such that $A=L(M^B)$.

\begin{definition} 
\begin{enumerate}
\item
{\bf \cite{sel:j:reductions-pselective,sel:j:ana}}  \quad
We say that a Turing machine $M$ is {\em positive\/} if
$$ (\forall A,B \subseteq \Sigma^*)
[ A \subseteq B \implies L(M^A) \subseteq L(M^B)].$$

\item 
{\bf \cite{sel:j:reductions-pselective}}  \quad
Let $A$ and $B$ be sets ($A,B\subseteq \Sigma^*$).
We say that $A$ {\em positive Turing reduces to\/} $B$ ($A \leq_{pos}^p B$) if
there is a polynomial-time positive Turing machine $M$ 
such that $A \leq_T^p B$ via $M$.



\end{enumerate}
\end{definition}

Since Selman's work, alternate definitions have been 
examined in some detail~\cite{hem-jai:j:pos}, and 
positive reductions have been seen to play a role in 
a number of places in complexity theory.
Most notably, Selman introduced them in the context of the 
P-selective sets, and to this day they continue to 
help in the investigation of those sets.
Positive reductions have also been used to characterize the 
class of languages that can be ``helped'' by unambiguous
sets~\cite{cai-hem-vys:b:promise}.

Henceforth, we will use ``positive Turing reductions'' as a
shorthand for ``polynomial-time positive Turing reductions.''
Selman's seminal work exactly pinpointed the power of positive Turing
reductions to NP, namely, the class of languages that positive Turing
reduce to NP is in fact NP itself.  The class of languages that Turing
reduce to NP, $\p^{\np}$, is a strictly larger class than this, unless
$\np = \conp$.  So, assuming that the polynomial hierarchy does not
collapse to NP, Turing reductions to NP are strictly more powerful
than positive Turing reductions to NP.

In this paper, we study the power of positive Turing reductions to 
DP\@.
DP was introduced by Papadimitriou and 
Yannakakis~\cite{pap-yan:j:dp}.  

\begin{definition} {\bf \cite{pap-yan:j:dp}} \quad
A set $C$ is in DP if there exist an NP set $A$ and 
a coNP set $B$ such that $C = A \cap B$.
\end{definition}

DP is the next
level beyond NP in the boolean 
hierarchy~\cite{cai-gun-har-hem-sew-wag-wec:j:bh1}, a structure
that has been used in contexts ranging from
approximation~\cite{cha:jtoappear:queries-approximations} to
query order~\cite{hem-hem-wec:j:query-order-bh}.
DP, by definition, is simply the class of languages 
that are the intersection of an NP and a coNP set, though
this class is quite robust and has many equivalent
definitions.
DP has natural complete problems (Graph Minimal 
Uncolorability~\cite{cai-mey:j:dp} 
and many others~\cite{wag:j:more-on-bh}), and plays 
a central role in the study of bounded access to NP,
due to its central role in the key normal form for 
the boolean hierarchy, which turns out to 
be exactly the finite unions of DP 
sets~\cite{cai-gun-har-hem-sew-wag-wec:j:bh1}.
DP also plays a role in 
the study of which sets are 
P-compressible~(\cite{gol-hem-kun:j:address}, see
also~\cite{wat:t:sparse}).

Clearly, $\np \subseteq \DP \subseteq \p^{\np}$.
Recall that Selman proved that positive Turing reductions to 
NP are surprisingly weak;  they yield just the NP sets.
In this paper, we prove that positive Turing reductions to 
DP are surprisingly strong;  they yield all the $\p^{\np}$ sets.
That is, they yield all the sets that can be computed via
Turing reductions to NP (equivalently, via Turing reductions
to DP.

We will note that our proof even establishes the same
level of power for 
$\p^{\np[1]}$, the class of languages computed by P
machines making at most one query to an NP oracle.

\section{On the Power of Positive Reductions to DP Sets}

We now prove our main result.   As is standard,
for any class $\cal C$ and any reducibility $r$, 
$${\rm R}_{r}^p({\cal C}) = \{ L \mid (\exists L' \in {\cal C})[L
\leq_r^p L']\},$$
that is, ${\rm R}_{r}^p({\cal C})$ is the class of sets that $r$-reduce to
sets in $\cal C$.

\begin{theorem}
${\rm R}_{pos}^p(\DP) = \p^{\np}$.
\end{theorem}

\medskip
\noindent {\bf Proof:} 
Clearly ${\rm R}_{pos}^p(\DP) \subseteq 
\p^{\DP} = \p^{\np}$.  So we have only to prove that 
${\rm R}_{pos}^p(\DP) \supseteq \p^{\np}$.

We will show that the standard 
$\p^{\np}$-complete problem
\OMS, the set of Boolean formulas whose lexicographically maximum
satisfying assignment is odd~\cite{kre:j:optimization},
is in ${\rm R}_{pos}^p(\DP)$.

In order to prove this, we will construct a polynomial-time positive
Turing machine $M$ such that
$L(M^{\SAT \oplus \overline{\SAT}}) = \OMS$.
For any sets $A$ and $B$, $A \oplus B$ denotes $\{x0 \  |\  x \in A\} \cup 
\{x1 \  | \ x\in B\}$.  The construction is reminiscent of the proof
that positive-truth-table reductions to tally sets
are as strong as truth-table reductions to tally
sets~\cite{buh-hem-lon:j:sparse}.

Define $M$ as follows.
$M$ will reject all strings that are not Boolean formulas.
Suppose $\phi$ is a Boolean formula on $n$ variables.
Without loss of generality, we assume that 
$x_1,\ldots,x_n$ are the variables of $\phi$.
$M$ on input $\phi$ works as follows:

Let $\phi_1 = \phi$.
For $i := 1$  to $n$: 

\begin{enumerate}
\item Query
$\phi_i[x_i := 1]0$ and $\phi_i[x_i := 1]1$.
\item If the answer to
$\phi_i[x_i := 1]0$ is ``yes'' and  the answer to $\phi_i[x_i := 1]1$
is ``no,'' then $\phi_{i+1} := \phi_i[x_i := 1]$.
If $i = n$, then accept.
\item If the answer to
$\phi_i[x_i := 1]0$ is ``no'' and  the answer to $\phi_i[x_i := 1]1$
is ``yes,'' then $\phi_{i+1} := \phi_i[x_i := 0]$.
If $i = n$, then reject.
\item If the answer to both queries is ``yes,'' then accept.
\item If the answer to both queries is ``no,'' then reject.
\end{enumerate}

Clearly, $M$ runs in polynomial time.
If we run $M$ on input $\phi$ with oracle
$\SAT \oplus \overline{\SAT}$, 
the answer to $\phi_i[x_i := 1]0$ is ``yes'' if and only if
$\phi_i[x_i := 1] \in \SAT$ and
the answer to $\phi[x_i := 1]1$ is ``yes'' if and only if
$\phi_i[x_i := 1] \not \in \SAT$.
So, for each iteration,  $M$ will be in case 2 or 3, so that
$M$ will accept $\phi$ if and only if $\phi$'s lexicographically
maximum satisfying assignment is odd. It follows that
$L(M^{\SAT \oplus \overline{\SAT}}) = \OMS$.
Since $\SAT \oplus \overline{\SAT} \in \DP$, it remains to show
that $M$ is positive.

Let $C,D,E,F$ be such that $C \oplus D \subseteq E \oplus F$.
Then $C \subseteq E$ and $D \subseteq F$. We will show that
for any string $x$, if $M^{C \oplus D}$ accepts $x$, then
$M^{E \oplus F}$ accepts $x$.  This is immediate for strings $x$ that
are not Boolean formulas, because they are rejected no matter what.
So, suppose that $x$ is a Boolean formula $\phi$ with
variables $x_1, \ldots, x_n$ and 
suppose for a contradiction that
$M^{C \oplus D}$ accepts $\phi$ and
$M^{E \oplus F}$ rejects $\phi$.

Let $i$ be the first iteration of the ``for'' loop such that 
$M^{C \oplus D}$ and $M^{E \oplus F}$ behave differently.

If $\phi_i[x_i := 1] \in E$ and $\phi_i[x_i := 1] \in F$
then $M^{E \oplus F}$ accepts, contradicting our assumption
that $\phi$ is rejected by $M^{E \oplus F}$.

If $\phi_i[x_i := 1] \not \in E$ 
and $\phi_i[x_i := 1] \not \in F$, then
$\phi_i[x_i := 1] \not \in C$ 
and $\phi_i[x_i := 1] \not \in D$,
and $M^{C \oplus D}$ rejects,
contradicting our assumption
that $\phi$ is accepted by $M^{C \oplus D}$.

So, it must be the case that
either $\phi_i[x_i := 1] \in E$  or
$\phi_i[x_i := 1] \in F$, but not both.
Since $M^{C \oplus D}$ and $M^{E \oplus F}$ behave differently at this stage,
and since $C \subseteq E$ and $D \subseteq F$,
it follows that $\phi_i[x_i := 1] \not \in C$ 
and $\phi_i[x_i := 1] \not \in D$.
But this implies that $M^{C \oplus D}$ rejects,
contradicting our assumption
that $\phi$ is accepted by $M^{C \oplus D}$.

This concludes the proof that $M$ is positive.
So, $M$ is a polynomial-time positive Turing machine such that
$L(M^{\SAT \oplus \overline{\SAT}}) = \OMS$.
Since $\SAT \oplus \overline{\SAT} \in \DP$, this implies
that $\OMS \in {\rm R}_{pos}^p(\DP)$, which implies that
${\rm R}_{pos}^p(\DP) \supseteq \p^{\np}$, since 
$\OMS$ is complete for $\p^{\np}$.
\quad~$\Box$

In fact, note that $\SAT \oplus \overline{\SAT}$ is 
not merely a DP set, but is  even in $\p^{\np[1]}$.  Thus, as an immediate
corollary to the proof, we can claim the following result

\begin{theorem}
${\rm R}_{pos}^p(\p^{\np[1]}) = \p^{\np}$.
\end{theorem}

Earlier, we mentioned that Selman's positive Turing reductions were
themselves inspired by the earlier notion of (polynomial-time)
positive truth-table reductions (see~\cite{lad-lyn-sel:j:com} for a
detailed formal definition of any notions used without 
definition in this paragraph).  The reader may wonder what the power of
positive truth-table reductions to DP is.  In fact, the answer to this
is already implicit in the existing literature.  Namely,
it is known that 
${\rm  R}^p_{disjunctive\hbox{-}truth\hbox{-}table}(\DP) = 
{\rm R}^p_{truth\hbox{-}table}
(\np)$~\cite{hem-hem-rot:j:raising-lower-bounds,bus-hay:j:tt}.
So, we may immediately 
conclude that ${\rm R}^p_{truth\hbox{-}table}(\np) = {\rm
  R}^p_{disjunctive\hbox{-}truth\hbox{-}table}(\DP) \subseteq 
{\rm R}^p_{positive\hbox{-}truth\hbox{-}table}(\DP)
\subseteq {\rm R}^p_{truth\hbox{-}table}(\np)$.  Thus, 
$$ {\rm R}^p_{positive\hbox{-}truth\hbox{-}table}(\DP) = {\rm R}^p_{
truth\hbox{-}table}(\np).$$ 
The class ${\rm R}^p_{truth\hbox{-}table}(\np)$, usually referred to as the 
``$\Theta_2^p$'' level of the polynomial 
hierarchy~(see \cite{wag:j:bounded}), has been
extensively studied,
and is widely believed to differ from 
$\p^{\np}$ (which obviously contains it).
However, 
``$\Theta_2^p = 
\p^{\np}$?''
remains a major open research question, 
and it is not even known whether this 
equality implies the collapse of the polynomial hierarchy.   From 
the main result of this paper, it is clear that 
``$\Theta_2^p = \p^{\np}$?'' can equivalently be stated as 
``${\rm R}^p_{positive\hbox{-}truth\hbox{-}table}(\np) = 
{\rm R}_{pos}^p(\np)$?''

{\samepage
\begin{center}
{\bf Acknowledgments}
\end{center}
\nopagebreak
\indent
For helpful conversations and suggestions, I am
very grateful to 
Lane Hemaspaandra, Harald Hempel, J\"org Rothe, and Gerd 
Wechsung.

}

\singlespacing
\bibliographystyle{alpha}

{\small

\newcommand{\etalchar}[1]{$^{#1}$}

}
\end{document}